\documentclass{article}
\usepackage{graphicx}

\usepackage[latin2]{inputenc}
\usepackage[T1]{fontenc}

\usepackage{amsmath}
\usepackage{amssymb}
\usepackage{enumerate}
\usepackage{amsmath}
\usepackage{amsthm}
\usepackage{amsfonts}
\usepackage{amssymb}
\usepackage{bbm}
\usepackage{color, soul}  
\usepackage{framed}
\usepackage{graphicx}
\usepackage{caption}
\usepackage{subcaption}
\usepackage[noend]{algorithmic}

\newcommand{\dd}{d}
\newcommand{\db}{db}
\newcommand{\dl}{dl}

\newcommand{\R}{\mathbb{R}}

\newcommand{\Ss}{\mathbb{S}}
\newcommand{\W}{\mathbb{W}}

\newcommand{\nt}{\noindent}

\newtheorem{asss}{Assumption}
\newtheorem*{asss21*}{Assumption 2.1}
\newtheorem*{asss22*}{Assumption 2.2}

\theoremstyle{remark}

\theoremstyle{plain}

\theoremstyle{remark}

\newtheorem*{remark*}{Remark}

\title{Exact sampling of diffusions with\\ a discontinuity in the drift}
\author{
	Omiros Papaspiliopoulos \thanks{Supported by MTM2012-37195 grant} \\
	ICREA $\&$ Department of Economics and Business,\\
	 Universitat Pompeu Fabra, Ramon Trias Fargas 25-27,\\
08005 Barcelona, Spain\\
	\and
Gareth O. Roberts \thanks{Supported by  EPSRC grants EP/D002060/1 and EP/K014463/1  } \\
	Department of Statistics\\
	University of Warwick  \\
	\and
	Kasia B. Taylor\thanks{Supported by grants: EPSRC EP/D002060/1, ERC 226488  and the Leverhulme Trust grant RPG-2013-270. The author would like to thank Dan Crisan for valuable comments related to the article.}\\
	Department of Statistics\\
	University of Warwick
	}

\date{}

\begin{document}

\maketitle

 \begin{abstract}
We introduce exact methods for the simulation of sample paths of one-dimensional
diffusions  with a discontinuity in the drift function.  Our procedures
require the simulation of finite-dimensional candidate draws from probability laws
related to those of Brownian motion and its local time and are based
on the principle of retrospective rejection sampling. A simple illustration is provided. 
\end{abstract}

\par{\cite{beskos2006retrospective} and \cite{beskos2008factorisation}  introduce a collection of efficient algorithms for the exact simulation of skeletons of diffusion processes. While
the methodology is intrinsically limited in the multivariate case to processes  which can be transformed into unit
volatility diffusions with drifts which can be written as the gradient of a potential, for one-dimensional non-explosive
diffusions the algorithm's 
applicability depends, more or less, only on  smoothness conditions on the diffusion and drift
coefficients. }
\par{However,  being able to simulate one-dimensional diffusions with discontinuous drifts is of considerable interest, not least
because this then allows us to tackle reflecting boundaries by suitable unfolding procedures. Therefore, in this paper we
focus on extending these  exact algorithms to the case of discontinuous drift. Namely, we consider solutions to one-dimensional SDEs of the form 
\begin{align}
\label{e:SDEfirst}
dX_{t}=\alpha(X_{t})dt+dB_{t}, \quad X_{0}=x, \quad t \in [0,T], \; T <\infty
\end{align}
where $B$ denotes  one-dimensional Brownian motion and $\alpha$ is a discontinuous function which satisfies  assumptions as specified later on.}
\par{These exact algorithms carry out rejection sampling in the
  diffusion trajectory space. The difficulty for discontinuous drift
  lies in the choice of suitable candidate measure for proposals in
  rejection sampling and in performing the acceptance/rejection step
  which reduces to sampling random variable with Bernoulli
  distribution with  unknown explicitly probability of success. We
  address these problems by suggesting a new methodology for sampling
  some conditional laws of Brownian motion and its local time. We
  construct a stochastic process to be used as proposal within
  rejection sampling in this context, which we call   ``local time
  tilted biased Wiener process''; this is to be contrasted with the simpler
  ``end-point tilted Wiener process'', which has been used when the drift is
  continuous. Overall, our approach is a natural evolution of the
  research program put forward in \cite{beskos2006retrospective} for
  simulation of diffusion sample paths using the Wiener-Poisson
  factorisation of diffusion measure together with retrospective
  rejection sampling principles.  The present work forms part of the
  doctoral thesis in \cite{taylor2015}.
  
  Concurrent to our work is that of \cite{etore2014exact}, who address the same fundamental problem as we do here. They follow a different approach from the one we undertake in this
  article,  one based on the limiting argument (with $n \to \infty$)
 applied to their earlier results for exact sampling of solutions to SDEs of the type $dX_{t}=\alpha(X_{t})dt+dB_{t}+\frac{1}{ n} dL_{t}$.
They then prove by indirect analytic arguments that their limiting algorithm does successfully simulate exactly from the SDE in (\ref{e:SDEfirst}). In contrast to that work, our paper offers a much simpler and more direct algorithm with a direct probabilistic interpretation as rejection sampling on path space. It is difficult to
be precise about the computational cost  comparison between the two methods,
although we estimate that our algorithms can be anything from 2-20 times quicker than that in 
\cite{etore2014exact}.

\section{Derivation of algorithms}
\label{sec:algs}

\par{Our aim here is to sample from $\mathbb{Q}$, the  measure  induced by the diffusion $X$ on $(C,\mathcal{C})$, the space of continuous functions on $[0,T]$ with the supremum norm and cylinder $\sigma$-algebra.  Note that the strong solution to (\ref{e:SDEfirst}) exists under mild conditions, namely if $\alpha$ is  bounded and measurable (see Theorem 4 in \cite{zvonkin1974transformation})}.
Denote by $\mathbb{W}$ the measure induced by Brownian motion started at $x$ on $(C,\mathcal{C})$. Under the following Assumption \ref{a:mgale}  we apply the Cameron-Martin-Girsanov theorem to obtain the Radon-Nikodym derivative of $\mathbb{Q}$ with respect to $\mathbb{W}$. 
\begin{asss}
\label{a:mgale}
The Cameron-Martin-Girsanov theorem holds and the Radon-Nikodym derivative $\frac{d \mathbb{Q}}{ d \mathbb{W}}$ is a true martingale:
\begin{align*}
\frac{d \mathbb{Q} }{ d \mathbb{W}}= \exp \Big \{\int_{0}^{T} \alpha(X_{t})dX_{t}-\frac{1}{ 2} \int_{0}^{T}\alpha^{2}(X_{t})dt \Big \}.
\end{align*}  
\end{asss}
The first step towards performing an acceptance/rejection step is the substitution of the stochastic integral $\int_{0}^{T} \alpha(X_{t})dX_{t}$. 
In existing Exact Algorithms for diffusions with sufficiently smooth coefficients
  (see e.g. \cite{beskos2005exact,beskos2006retrospective}) this is done by application of It\^{o}'s lemma to $A(X_t)$ where
   $A:=\int_{0}^{u} \alpha(y)dy$. 
   The discontinuity of $\alpha $ precludes
 proceeding in the same way, 
 although
  we can generalise the approach by
 appealing to the generalised It\^{o} formula. 
\begin{asss}
\label{as:karatzas}
Let $r_1<r_2<\cdots<r_n$ be real numbers, and define $D=\{r_1,r_2,\cdots,r_n\}$. Assume that the drift function $\alpha:\R \rightarrow \R$ is continuous on $\R \smallsetminus D$ and $\alpha'$ exists and is continuous on $\R \smallsetminus D$, and the limits
\begin{align*}
 \alpha'(r_k +):=\lim_{x \downarrow r_k } \alpha'(x) \quad \text{ and } \quad  \alpha'(r_k -):=\lim_{x \uparrow r_k } \alpha'(x), \quad k \in \{1,\cdots,n\}
\end{align*}
exist and are finite.
\end{asss}
\par{Under  Assumption \ref{as:karatzas} $A$ is  the difference of two convex functions and 
\begin{align*}
A(X_t)&=A(x)+\int_0^t \alpha(X_s)dX_s+\frac{1}{ 2} \int_0^t \alpha'(X_s)ds \\
&+\frac{1}{ 2}\sum_{k=1}^n L_{t}^{r_k} [ \alpha(r_k +)- \alpha(r_k -)], \;
 \text{a.s.  for }  t \in [0,T].
\end{align*} 
\par{The algorithms that we present address the problem in the case where
the drift function $\alpha$ has one point of discontinuity which
without loss of generality can be assumed to be at $0$, and denote the
discontinuity jump by  
\[\theta=\frac{1}{ 2}( \alpha(0 +)- \alpha(0 -))\,.
\]
Then the Radon-Nikodym derivative of $\mathbb{Q}$ with respect to $ \mathbb{W}$ is equal to
\begin{align}
\label{e:key}
\frac{d \mathbb{Q} }{ d \mathbb{W}}  &= \exp \big \{A(X_{T})-A(x) -\frac{1}{ 2} \int_{0}^{T}  (  \alpha'(X_{t})+\alpha^{2}(X_{t}))dt 
-\theta L^0_{T}(X)  \big \} \;\text{a.s.}
\end{align}
\par{The main question here is how to use (\ref{e:key}) for simulation.
 If (\ref{e:key}) is used in rejection sampling it needs to be uniformly bounded above. 
\begin{asss}
\label{a:bounded}
There exist constants $\kappa$ and $M$ with
$$
-\infty < \kappa\le \frac{1}{ 2}(
\alpha^{2}(u)+\alpha'(u))\mathbbm{1}_{u  \neq 0}\le \kappa+M.
$$
\end{asss}
\noindent
On the basis of this assumption, we define 
\[
\varphi(u)= \frac{1}{ 2}(
\alpha^{2}(u)+\alpha'(u))\mathbbm{1}_{u \neq 0}-\kappa\,.
\]
\par{ The main problem here is that multiplicative term $ \exp
  \{A(X_{T})-\theta L^0_T(X)\}$  may be unbounded. 
We address this issue  by biasing - using exponential tilting -  the
dominating measure $\mathbb{W}$, by these terms. 
 Thus we define $\mathbb{S}$ on $(C,\mathcal{C})$ with paths starting  at $x$
and satisfying}
\begin{align}
\label{e:S}
\frac{d \Ss }{ d \W }\propto \mathbbm{1}_{\{L_{T}(X)>0\}}(X)   \frac{g(X_{T}, L_{T}(X)) }{ f^{x}_{T}(X_{T}, L_{T}(X))}+\mathbbm{1}_{\{L_{T}(X)=0\}} (X)\frac{g_{*}(X_{T})  }{ f_{*,T}^{x}(X_{T}) }
\end{align}
with the following definitions
\begin{align}
\label{e:g1}
&g(b,l) :=c_{g} f_{T}^{x}(b,l)e^{A(b)-l \theta}\,,\quad
 \text{ for }l>0 \\
 \label{e:g2}
&g_{*}(b) :=c_{g} f_{*,T}^{x}(b)
e^{A(b)} \\ 
& \text{such that} \quad \int_{\R}\int_{ (0,\infty)} g(b,l)  \dd l \dd b+\int_{\R} g_{*}(b) \dd b=1. \nonumber
\end{align}
Above,  $f$ and $f_{*}$ describe joint law of Brownian motion and its local time at level zero (see e.g. \cite{borodin2002handbook})
\begin{align}
\label{e:f1}
 f_{s}^{x}(b,l)\db \; \dl&:=P(B_{s}\in \db,L_{s}\in \dl | B_{0}=x)
 = (  s \sqrt{2\pi s})^{-1}(l+|b|+|x|) \nonumber  \\
 & \times \exp \Big \{  -\frac{  (l+|b|+|x|)^{2} }{ 2s}\Big \}\;  \db\; \dl, \quad  \text{ for } l >0
\end{align}
 If  $x>0$ and $b>0$ (or $x<0$ and $b<0$) then
\begin{align}
\label{e:f2}
 f_{*,s}^{x}(b)\db &:= P(B_{s}\in \db,L_{s}=0 | B_{0}=x)=\phi_{x,s}(b)\big (1-e^{-\frac{2bx}{ s}}\big)db.
\end{align}
The above definitions of $g$ and $g_{*}$ rely on Assumption \ref{as:g} below. 
 \begin{asss}
 \label{as:g}
$\int_{\R}\int_{(0,\infty) }f_{T}^{x}(b,l)e^{A(b)-l \theta} dl db <\infty$ and\\
 $\int_{R} f_{*,T}^{x}(b) e^{A(b)} db <\infty$
\end{asss}
A fairly direct calculation then yields, 
\begin{align*}
\frac{d \mathbb{Q}}{ d \mathbb{S}}  \propto \exp \Big \{  -
\int_{0}^{T} \varphi(X_t) \; dt \Big \} \;\text{a.s.}
\end{align*}
 \par{\cite{beskos2006retrospective} show that for Radon-Nikodym
 derivatives of this form for $\phi>0$ and bounded, and provided
 finite-dimensional distributions of $\mathbb{S}$ can be simulated
 exactly, exact simulation of sample paths from $\mathbb{Q}$ is
 feasible using retrospective rejection sampling using auxiliary
 Poisson processes.  We present now   Algorithm 1   which requires
 Assumptions \ref{a:mgale}, \ref{as:karatzas} (with $n=1$),
 \ref{a:bounded} and   \ref{as:g}. We denote by
 $\{0=t^{0},t^{1},\cdots,t^{n},t^{n+1}=T\}$ the time instances at
 which we want to sample the diffusion. Recall that  $X_{0}=x$. The
 following algorithm returns values of the diffusion $X$ together with
 its local time at collection of  chosen and random time points. }

\noindent\makebox[\linewidth]{\rule{\textwidth}{0.4pt}}

\noindent
\textbf{Exact Algorithm for drift with discontinuity} \\
\noindent\makebox[\linewidth]{\rule{\textwidth}{0.4pt}}

\begin{description}
 \item[I] Generate $(X_T, L_T)$ according to law given by (\ref{e:g1}) and (\ref{e:g2})
\item[II] Sample an auxiliary  Poisson process $\Psi$ on $[0,T]$ with a rate parameter $M$ to get $(\tau_{1}, \cdots, \tau_{k})$ and then $\psi_{i} \sim U(0,M)$ i.i.d. for $i=1,\cdots, k$
\item[III] Generate $X$ and $L$ at times $(\tau_{1},\tau_{2}, \cdots, \tau_{k})$  conditioned on values at $0$ and $T$ (see Sections \ref{sec:restricted:BB} and \ref{sec:double:bridge})

\item[IV] Use $\Psi$ to perform  rejection sampling:
		\begin{description}
		\item[(i)] Compute $\varphi(X_{\tau_{i}})$ for $i \in \{1,\cdots, k\}$
		\item[(ii)] If  $\varphi(X_{\tau_{i}}) < \psi_{\tau_{i}}$ for each $i \in \{1,\cdots, k\}$ then proceed to (V)
		otherwise start again at (I)
	\end{description}
	\item[V] Generate $X$ jointly with $L$ at times $(t^{1},\cdots,t^{n})$ conditioned on values at $(0,\tau_{1},\tau_{2}, \cdots, \tau_{k},T)$
(Sections \ref{sec:restricted:BB} and \ref{sec:double:bridge}).
\item[VI] Return
$
(0,x,0),(t_{1},X_{t_{1}},L_{t_{1}}),\cdots,(t_{k+n},X_{t_{k+n}},L_{t_{k+n}}),(T,X_{T},L_{T})
$
\end{description}
\noindent\makebox[\linewidth]{\rule{\textwidth}{0.4pt}}

\bigskip

The practical and probabilistic challenges with this algorithm are
contained in Step I, which requires simulation from a  bivariate
distribution on the endpoint of path and its accumulated local time at
0, and Step III which requires finite-dimensional
distributions of Brownian motion and its local time. In the rest of
this section we address the first challenge and in Section
\ref{sec:bridges} the second challenge. However, as we show below solving
the second problem can also provide alternative and more efficient
solutions to the first, given an additional assumption.

\subsection{Simulating the end-point distribution}

We first provide some further insight on the exponential tilting employed
in the definition of $\mathbb{S}$. Note that we are biasing the joint
distribution of $(X_T,L_T(X))$ under the Wiener measure, with the terms
$e^{A(X_T)}$ and  $e^{-\theta L_T(X)}$. 
This observation
generates our first method for simulating from $\mathbb{S}$, when  the
discontinuity jump is positive, i.e., if $\theta>0$. We 
decompose the law of $(X_T,L_T(X))$ as the marginal of $X_T$ and the
conditional of $L_T(X) | X_T$. The simulation of $X_T$ in this case is
done according to 
\begin{align}
\label{eq:def:h}
h(u) \; \propto \; \exp\{ A( u) \}\phi_{x,T} (u) 
\end{align}
where $\phi_{\mu,\sigma^{2}}$  denotes the density
($\Phi_{\mu,\sigma^{2}}$ denotes respectively cumulative distribution
function) of the normal distribution with mean $\mu$ and variance
$\sigma^{2}$. This step is in fact common to all exact algorithms for
diffusions. Conditionally on $X_T$ drawn according to this scheme,
$L_T|X_T$ is proposed according to the conditional law of local time
given point evaluation under the Wiener measure, as described in  Section
\ref{sec:bridges}. Any value produced in this way is then accepted
with probability $e^{-\theta L_T(X)}$. An accepted pair $(X_T,L_T(X))$
produced by this procedure is drawn from $\mathbb{S}$. 

When the jump is negative, $\theta <0$, the above procedure cannot
work and the biasing due to the local time has to be dealt in a
slightly more involved way, albeit still by rejection sampling. We
will simulate $(X_T,L_T(X))$ jointly in this case, and find it useful to use mixture distribution consisting of six mixture components appropriately chosen to dominate  either the tails or the central part of probability distribution given by $g$ and $g_{*}$ in case $(b\geq 0,l >0), (b<0,l>0)$ or $(b> 0,l=0)$ (or $(b< 0,l=0)$ if $x<0$). 
Suppose $\xi_{1},\xi_{3}>0$ and $\xi_{2},\xi_{4}<0$. Let $h_{i}:\R \times (0,\infty) \to \R$ ($i=1,\dots,4$) and $h_{i}:\R \to \R $ ($i=5,\dots,8$)  be probability distributions such that  
\begin{align*}
h_{1}>0 \; \text{ iff } \; (b,l) \in [0,\xi_{1}]\times (0,\infty)\quad  & \quad h_{2}>0 \; \text{ iff } \; (b,l) \in  (\xi_{1},\infty)\times (0,\infty)\\
h_{3}>0 \; \text{ iff } \; (b,l) \in  [\xi_{2},0)\times (0,\infty)\quad  & \quad h_{4}>0 \; \text{ iff } \; (b,l) \in  (-\infty,\xi_{2})\times (0,\infty)\\
h_{5}>0 \; \text{ iff } \; b \in [0,\xi_{3}]  \quad  & \quad h_{6}>0 \; \text{ iff } \; b \in (\xi_{3},\infty)\\
h_{7}>0 \; \text{ iff } \; b \in[\xi_{4},0] \quad  & \quad h_{8}>0 \; \text{ iff } \; b \in (-\infty,\xi_{4})
\end{align*}
We use $h_{i}$'s with $i \in \{1,\cdots,6\}$ when $x>0$, with $i \in \{1,2,3,4, 7,8\}$ when $x<0$ and with $i \in \{1,\cdots,4\}$ when $x=0$. Assume  for now that $x>0$. Chose $K$ such that $\frac{g(b,l) }{ \frac{1 }{ 6} h_{i}(b,l) } <K$ whenever $h_{i}>0$ for $ i=1,\cdots 4$ and $\frac{g_{*}(b) }{ \frac{1 }{ 6} h_{i}(b) } <K$ whenever $h_{i}>0$ for $ i=5,6$ 
  then the procedure of sampling the candidate end pair $(b,l)$ is as
  described below. 

\noindent\makebox[\linewidth]{\rule{\textwidth}{0.4pt}}

\noindent
\textbf{Endpoint rejection sampling} \\
\noindent\makebox[\linewidth]{\rule{\textwidth}{0.4pt}}

\begin{algorithmic}
\REPEAT
	\STATE  $U_{1} \sim$ DiscreteUniform on $\{1,2,\cdots,6\}$
	\STATE  $U_{2} \sim$ Unif$(0,1)$
	\FOR{$i=1$ to $4$}
	\vspace{-0.5cm}
		\STATE \IF {$u_{1}=i$}
		   \STATE  $(b,l) \sim h_{i}$
		   	\vspace{-0.25cm}
		    \STATE  \IF {$u_{2} \leq \frac{g(b,l)}{ K \frac{1 }{ 6}h_{i}(b,l) } $}
		    	\STATE accept $(b,l)$
			\ENDIF
			\ENDIF
	\ENDFOR
	\FOR{$i=5$ to $6$}
		\vspace{-0.5cm}
		\STATE \IF {$u_{1}=i$}
		   \STATE  $b \sim h_{i}$
		   \vspace{-0.25cm}
		    \STATE  \IF {$u_{2} \leq \frac{g_{*}(b) }{ K \frac{1 }{ 6}h_{i}(b) } $}
		    	\STATE accept $b$, set $l=0$
			\ENDIF
			\ENDIF
	\ENDFOR	
\UNTIL  {$b$ accepted}
\RETURN $(b,l)$
\end{algorithmic}
\noindent\makebox[\linewidth]{\rule{\textwidth}{0.4pt}}

\section{Sampling procedures for Brownian motion and its local time}
\label{sec:bridges}

\par{In this section we discuss sampling from the conditional laws
 $\mathcal{L}(L_{s_{2}} | B_{s_{1}}, \hspace{-0.05cmB_{s_{2}}, \hspace{-0.05cm}L_{s_{1}})$ and  $\mathcal{L}(B_{s_{2}},L_{s_{2}}|B_{s_{1}},B_{s_{3}},L_{s_{1}},L_{s_{3}})	$ where $0 \leq s_{1} <s_{2}<s_{3}$.
Note that  $(B,L)$ is a Markov process, with increments whose
distribution only depends on the first coordinate of the process,
which  facilitates simulation of finite-dimensional distributions of
the process as well as finite-dimensional distributions of the process
conditionally on its endpoints. 
Further we observe that as far as simulating
skeletons of the process conditionally on a starting point, it suffices
to describe how to generate from  $\mathcal{L}(L_{s_{2}} |
B_{s_{1}},B_{s_{2}}, L_{s_{1}})$, which we do in Section
\ref{sec:one:end}. Simulating finite-dimensional distributions
conditionally on past and future endpoints requires simulation from 
 $\mathcal{L}(B_{s_{2}},L_{s_{2}}|B_{s_{1}},B_{s_{3}},L_{s_{1}},L_{s_{3}})$;
 this problem we tackle in two steps, first in the simpler case where
 $L_{s_{1}}=L_{s_{3}}$, in which case $L_{s_2}$ is also
 equal to those values, and then for the general case where the local
 time has changed between the two endpoints. }
 
\subsection{Sampling from $\mathcal{L}(L_{s_{2}} | B_{s_{1}},B_{s_{2}}, L_{s_{1}})$}
\label{sec:one:end}

\par{We compute  probability that the local time
  does not increase on $[s_{1},s_{2}]$.  Assuming $b_{1} <0$ and $b_{2}<0$ (or $b_{1}>0$ and $b_{2}>0$)
\begin{align*}
P(L_{s_{2}} = l_{1}| B_{s_{1}}=b_{1},B_{s_{2}}=b_{2},L_{s_{1}}=l_{1})& \; = \; \frac{f^{b_{1}}_{*,s_{2}-s_{1}}(b_{2})               }{ \phi_{b_{1},s_{2}-s_{1}}(b_{2})         }  
= 1- e^{\frac{- 2b_{1}b_{2} }{ s_{2}-s_{1}}}.
\end{align*}
Now suppose that  $l_{2}>l_{1}$
\begin{align*}
&P(L_{s_{2}} \in \dd l_{2}| B_{s_{1}}=b_{1},B_{s_{2}}=b_{2},L_{s_{1}}=l_{1})
=\frac{f^{b_{1}}_{s_{2}-s_{1}}(b_{2},l_{2}-l_{1})                    }{ \phi_{b_{1},s_{2}-s_{1}}(b_{2})       }   \dd l_{2}\\
&=\frac{  1 }{  s_{2}-s_{1}}((l_{2}-l_{1})+|b_{2}|+|b_{1}|)e^{  \frac{ - ((l_{2}-l_{1})+|b_{2}|+|b_{1}|)^{2} }{2(s_{2}-s_{1})}}e^{  \frac{ (b_{2}-b_{1})^{2} }{ 2(s_{2}-s_{1})}} \;   
                    \dd l_{2} \\
&\text{and further by substituting } \;  l:=(l_{2}-l_{1})+|b_{2}|+|b_{1}| \\ 
&\propto \frac{  l }{  s_{2}-s_{1}}e^{  -\frac{  l^{2} }{ 2(s_{2}-s_{1})}} \dd l. 
\end{align*}}
\begin{framed}

\nt
\textbf{Sampling from $\mathcal{L}(L_{s_{2}} | B_{s_{1}},B_{s_{2}}, L_{s_{1}})$ with $B_{s_{1}}=b_{1},B_{s_{2}}=b_{2}, L_{s_{1}}=l_{1}$}\\

\nt
\begin{enumerate}
\item If $b_{1}<0< b_{2}$ or $b_{1}> 0> b_{2}$ proceed to (2.) \\
              Otherwise
              \subitem        Sample $U \sim U(0,1)$
               \subitem    If $u \leq  1- e^{\frac{- 2b_{1}b_{2}  }{ s_{2}-s_{1}}}
$ set $l_{2}=l_{1}$ and finish here
               \subitem     otherwise proceed to (2.)
\item  Sample $Z \sim U\big(1-e^{-\frac{(|b_{2}|+|b_{1}|)^{2} }{ 2 (s_{2}-s_{1})}},1\big)$\\
	Set $y=\sqrt{-2(s_{2}-s_{1})\ln(1-z)}$

	 Set $l_{2}=y+l_{1}-|b_{2}|-|b_{1}|$
	 \end{enumerate}
\nt

\end{framed}

\bigskip

\subsection{Sampling from $\mathcal{L}(B_{s_{2}},L_{s_{2}}|B_{s_{1}},B_{s_{3}},L_{s_{1}},L_{s_{3}})	$ where $L_{s_{1}}=L_{s_{3}}$ }
\label{sec:restricted:BB}

In this and following Section \ref{sec:double:bridge} we are
interested in conditioning Brownian motion and its local time on both
past and future values. Here we consider $L_{s_{1}}=L_{s_{3}}$ which implies that  $B_{s} \neq 0$  for all $s \in [s_{1},s_{3}]$  a.s.. Also  here $L_{s_{2}}$ is trivially equal to $L_{s_{1}}$. 
Throughout this section we suppose $b_{1}>0$, $b_{2}>0$ and $b_{3}>0$;
the case $b_{1}<0$, $b_{2}<0$ and $b_{3}<0$ can be treated completely symmetrically.

Using Bayes' theorem 
  obtain
\begin{align}
\label{eq:mu:sigma}
\nu_{1}(db_{2})&:=P(B_{s_{2}} \in d b_{2}|B_{s_{1}}=b_{1},B_{s_{3}}=b_{3},L_{s_{1}}=l_{1},L_{s_{3}}=l_{1})  \nonumber\\
&=\frac{f^{b_{1}}_{*,s_{2}-s_{1}}(b_{2})        f^{b_{2}}_{*,s_{3}-s_{2}}(b_{3}) }{ f^{b_{1}}_{*,s_{3}-s_{1}}(b_{3})}db_{2}  \nonumber\\
&= \phi_{  \mu, \sigma^{2}}(b_{2})  \Big(1-e^{\frac{-2b_{2}b_{1}}{ s_{2}-s_{1}}}\Big)  
  \Big(1-e^{\frac{-2b_{3}b_{2}}{ s_{3}-s_{2}}}\Big)    \Big(1-e^{\frac{-2b_{3}b_{1}}{ s_{3}-s_{1}}}\Big)^{-1}
  e^{\frac{-2b_{3}b_{1}}{ s_{3}-s_{1}}}
  db_{2}  \nonumber\\
\text{where }\mu& := \frac{ b_{1}(s_{3}-s_{2})+b_{3}(s_{2}-s_{1})   }{ s_{3}-s_{1}    } \quad \sigma^{2}:={(s_{2}-s_{1})(s_{3}-s_{2})  \over s_{3}-s_{1} }
\end{align}
Next we introduce measure $\nu_{2}$ which will be useful for an auxiliary rejection sampling. 
\begin{align*}
\nu_{2}(db_{2})&:=P(B_{s_{2}} \in d b_{2}|B_{s_{1}}=b_{1},B_{s_{3}}=b_{3})\\
& \quad  \times (P(B_{s_{2}}>0|B_{s_{1}}=b_{1},B_{s_{3}}=b_{3}, b_{1}>0,b_{3}>0))^{-1}\\
 {d\nu_{1} \over d\nu_{2}} &= \Big(1-e^{{-2b_{2}b_{1}\over s_{2}-s_{1}}}\Big)\Big(1-e^{{-2b_{3}b_{2}\over s_{3}-s_{2}}}\Big)
 e^{{-2b_{3}b_{1}\over s_{3}-s_{1}}}  
 \Big(1-e^{{-2b_{3}b_{1}\over s_{3}-s_{1}}}\Big)^{-1} \Big (1- \Phi_{\mu,\sigma^{2}}(0)\Big).
\end{align*}
\begin{oframed}
\nt
\textbf{Sampling from $\mathcal{L}(B_{s_{2}}|B_{s_{1}},B_{s_{3}},L_{s_{1}},L_{s_{3}})	$ where $L_{s_{1}}=L_{s_{3}}$}\\

\nt
\begin{enumerate}
\item Sample $Z \sim $ truncated normal distribution $N(\mu, \sigma^2)$ on $(0,\infty)$  where $\mu$ and $\sigma^{2}$ as in (\ref{eq:mu:sigma})
\item Sample $U \sim U(0,1)$ 
\item  If $u <   {d\nu_{1} \over d\nu_{2}} $ 
 then set $b_{2}=z$ otherwise start again at (1.)
             	 \end{enumerate}

\end{oframed}

\bigskip
\subsection{Sampling from $\mathcal{L}(B_{s_{2}},L_{s_{2}}|B_{s_{1}},B_{s_{3}},L_{s_{1}},L_{s_{3}})$ where $L_{s_{1}} \neq L_{s_{3}}$}
\label{sec:double:bridge}

\bigskip
In this Section we concentrate on sampling Brownian motion and its local time at $s_{2}$  when $L_{s_{1}}=l_{1} \neq L_{s_{3}}=l_{3}$.

 We need to consider three cases, 
namely when local time stays constant over $[s_{1},s_{2}]$,  over $[s_{2},s_{3}]$  and the last one when is not constant over any of these intervals.\\

\begin{enumerate}
\item Suppose $l_{1} = l_{2}$ but $l_{1}\neq l_{3}$. Note that here we only consider $b_{1}>0$ and $b_{2}>0$ (or $b_{1}<0$ and $b_{2}<0$). Define $\xi_{1}(b_{2},l_{1})$ and $p_{1}$ as follows
\begin{align*}
&P(B_{s_{2}}\in \dd b_{2},L_{s_{2}} =l_{1}| B_{s_{1}}=b_{1},B_{s_{3}}=b_{3},L_{s_{1}}=l_{1},L_{s_{3}}=l_{3}) =: \xi_{1}(b_{2},l_{1}) db_{2}\\
&p_{1}:=\int_{-\infty}^{\infty}\xi_{1}(b_{2},l_{1}) db_{2}=P(L_{s_{2}} =l_{1}| B_{s_{1}}=b_{1},B_{s_{3}}=b_{3},L_{s_{1}}=l_{1},L_{s_{3}}=l_{3}) 
\end{align*}
The upper or the lower limit of integration above, depending if $b_{1}<0$ or $b_{1}>0$ respectively, can be changed to $0$.
\item Suppose $l_{2} = l_{3}$ but $l_{1}\neq l_{3}$. Note that here we only consider $b_{3}>0$ and $b_{2}>0$ (or $b_{3}<0$ and $b_{1}<0$). Define $\xi_{3}(b_{2},l_{3})$ and $p_{3}$ as follows
\begin{align*}
&P(B_{s_{2}}\in \dd b_{2},L_{s_{2}} =l_{3}| B_{s_{1}}=b_{1},B_{s_{3}}=b_{3},L_{s_{1}}=l_{1},L_{s_{3}}=l_{3}) =: \xi_{3}(b_{2},l_{3}) db_{2}\\
&p_{3}:=\int_{-\infty}^{\infty}\xi_{3}(b_{2},l_{3}) db_{2}=P(L_{s_{2}} =l_{3}| B_{s_{1}}=b_{1},B_{s_{3}}=b_{3},L_{s_{1}}=l_{1},L_{s_{3}}=l_{3}) 
\end{align*}
The upper or the lower limit of integration above, depending if $b_{3}<0$ or $b_{3}>0$ respectively, can be changed to $0$.
\item Suppose $l_{2} \in (l_{1},l_{3})$ and $l_{1}\neq l_{3}$. Define $\xi_{2}(b_{2},l_{2})$ and $p_{2}$ as follows
\begin{align*}
&P(B_{s_{2}}\in \dd b_{2}, \hspace{-0.08cm}L_{s_{2}} \in dl_{2}| B_{s_{1}}=b_{1}, \hspace{-0.08cm} B_{s_{3}}=b_{3},  \hspace{-0.08cm} L_{s_{1}}=l_{1},  \hspace{-0.08cm} L_{s_{3}}=l_{3}) =: \xi_{2}(b_{2},l_{2}) db_{2}dl_{2}\\
&p_{2}:=\int_{(l_{1},l_{3})}\int_{\R}\xi_{2}(b_{2},l_{2})db_{2}dl_{2} \\
&=P(L_{s_{2}}  \in (l_{1},l_{3})| B_{s_{1}}=b_{1},B_{s_{3}}=b_{3},L_{s_{1}}=l_{1},L_{s_{3}}=l_{3}) 
\end{align*}
\end{enumerate}
We introduced $p_{1},p_{2}$ and $p_{3}$ (where $p_{1}+p_{2}+p_{3}=1$) so that we can use it to split simulation into two steps. First we  determine the case and then conditioned on the case we sample the value of $B_{s_{2}}$ (or  in case (3.) of both: $B_{s_{2}}$ and $L_{s_{2}}$).
Observe that
\begin{align*}
& \xi_{2}(b_{2},l_{2}) ={f^{b_{1}}_{s_{2}-s_{1}}(b_{2},l_{2}-l_{1})             f^{b_{2}}_{s_{3}-s_{2}}(b_{3},l_{3}-l_{2})              \over f^{b_{1}}_{s_{3}-s_{1}}(b_{3},l_{3}-l_{1})    } \nonumber\\
&=c(l_{2}-l_{1}+|b_{2}|+|b_{1}|)e^{  -{  (l_{2}-l_{1}+|b_{2}|+|b_{1}|)^{2} \over 2(s_{2}-s_{1})}}       
 (l_{3}-l_{2}+|b_{3}|+|b_{2}|)e^{- { (l_{3}-l_{2}+|b_{3}|+|b_{2}|)^{2} \over 2(s_{3}-s_{2})   }} \\
&\text{where }  c:= 0.5 \pi^{-1}(s_{2}-s_{1})^{-{3\over2}}(  (s_{3}-s_{2})^{-{3\over2}}(f^{b_{1}}_{s_{3}-s_{1}}(b_{3},l_{3}-l_{1})          )^{-1}
 \end{align*}
Note in the above formula for $ \xi_{2}(b_{2},l_{2})$ \hspace{-0.15cm}  the symmetry in $b_{2}$ about  $0$,   \hspace{-0.15cm} i.e. $ \xi_{2}(y,l_{2})=$\\ $= \xi_{2}(-y,l_{2})$ for $y \in \R$ and $l_{2}\in(l_{1},l_{3})$. Assume that  $b_{2}>0$ then by substituting $u$ and $v$ defined as follows
\begin{align}
\label{eq:uv}
u:=l_{2}-l_{1}+|b_{2}|+|b_{1}| \qquad 
v:=l_{3}-l_{2}+|b_{3}|+|b_{2}|
\end{align}
we further have
\begin{align}
\label{eq:xi}
 \xi_{2}(b_{2},l_{2}) \propto ue^{  -{  u^{2} \over 2(s_{2}-s_{1})}}        ve^{- { v^{2} \over 2(s_{3}-s_{2})   }}.    
 \end{align}

Recall that  $\mathcal{L}(B_{s_{2}},L_{s_{2}}|B_{s_{1}},B_{s_{3}},L_{s_{1}},L_{s_{3}})$ is a.s. equal to $0$ outside $\R \times [l_{1}, l_{3}]$. Under the linear  transformation $(b_{2},l_{2}) \to (u,v)$ given by (\ref{eq:uv})  we have that the region $R_{1}:=[0,\infty)\times [l_{1}, l_{3}]$ is mapped onto the region $R_{2}$ bounded by the following lines:
\begin{align*}
v& =u+l_{1}-l_{3}-|b_{1}|+|b_{3}| \\
v&  =u-(l_{1}-l_{3}) -|b_{1}|+|b_{3}|\\
v& =-u-l_{1}+l_{3}+|b_{1}|+|b_{3}|.
\end{align*}
Observe that form of (\ref{eq:xi}) allows us to use sampling of two independent random variables with Rayleigh distribution  but with different scale parameters. However it is important to remember that this distribution needs to be truncated to region $R_{2}$. 

\begin{oframed}
\nt
\textbf{Sampling from $\mathcal{L}(B_{s_{2}},L_{s_{2}}|B_{s_{1}},B_{s_{3}},L_{s_{1}},L_{s_{3}})$ where $L_{s_{1}} \neq L_{s_{3}}$} \\

\nt
	\begin{description}
	\item{1.}  Sample $U \sim U(0,1)$
	\item{2.} Compute $p_{1}$. If $u>p_{1}$ proceed to (3.) otherwise
		\subitem Set $l_{2}=l_{1}$
		\subitem Sample $Z \sim h(z) \propto \xi_{1}(z,l_{1})$
		\subitem Set $b_{2}=z$ and finish here
	\item{3.} Compute $p_{3}$. If $z_{1}>p_{1}+p_{3}$  proceed to (4.) otherwise
		\subitem Set $l_{2}=l_{3}$
		\subitem Sample $Z \sim h(z) \propto \xi_{3}(z,l_{3})$
		\subitem Set $b_{2}=z$ and finish here
	\item{4.} Sample $(U,V)\sim h(u,v) \propto \mathbbm{1}_{(u,v) \in R_{2}}(u,v)ue^{  -{  u^{2} \over 2(s_{2}-s_{1})}}        ve^{- { v^{2} \over 2(s_{3}-s_{2})   }}   $ 
	\item \hspace{0.5cm} Set $l_{2}= { u-v+l_{3}+l_{1}-|b_{1}| +|b_{3}|  \over 2}   $ 
	\item \hspace{0.5cm}Sample $Y \sim \text{Ber}(0.5)$
	\item  \hspace{0.5cm}If $y=1$ then set  $b_{2}=   {   u+v-l_{3}+l_{1}-|b_{1}| -|b_{3}| \over 2}$
	\item  \hspace{0.5cm}otherwise set  $b_{2}= - {   u+v-l_{3}+l_{1}-|b_{1}| -|b_{3}| \over 2}  $.
	\end{description}

\end{oframed}
Explicit formulas for $p_{1}$ and $p_{3}$
 can be found in the Appendix.

\section{Examples}

\par{In this Section we present simple examples of numerical simulation of diffusions with discontinuous drift, namely satisfying SDEs: (1) $dX_{t}=a_{i}dt+dB_{t}$ and (2) $dX_{t}=sin(X_{t}-\theta_{i})dt+dB_{t}$ with $a_{1}, \theta_{1}$ if $X_{t}\geq 0$ and $a_{2}, \theta_{2}$ if $X_{t}<0$. }
\par{We produce 100000 observations of diffusions at time $T=1$  applying our exact methods and we use them for kernel density estimation. Our method is substantilly quicker than using the Euler-Maruyama scheme with $\Delta t=0.0001$ (as used for diagrams below) or even $\Delta t=0.001$. Moreover coarser discretisation leave the Euler-Maruyama competitive on running time, but with appreciable bias. In each example we set  $X_{0}=0$ to observe the effect of the discontinuity in the drift. }

\par{The computing time for  implementation of these algorithms on an Apple MacBook Air computer 1.86 GHz Intel Core 2 Duo were around 500s when using mixture distribution to produce candidate $(X_{T},L_{T})$ and 50s when two step procedure (first sampling $X_{T}$ then $L_{T}|X_{T}$) was applied, figures which compare favourably with \cite{etore2014exact} which reported running times of over 1000s on a similar example. All code was implemented in R, and it is considered that the algorithm using mixture distribution could be much more efficient with optimised code.}

\begin{figure*}[h]
	\begin{subfigure}[t]{5.5cm}
		\centering
		\includegraphics[width=\textwidth]{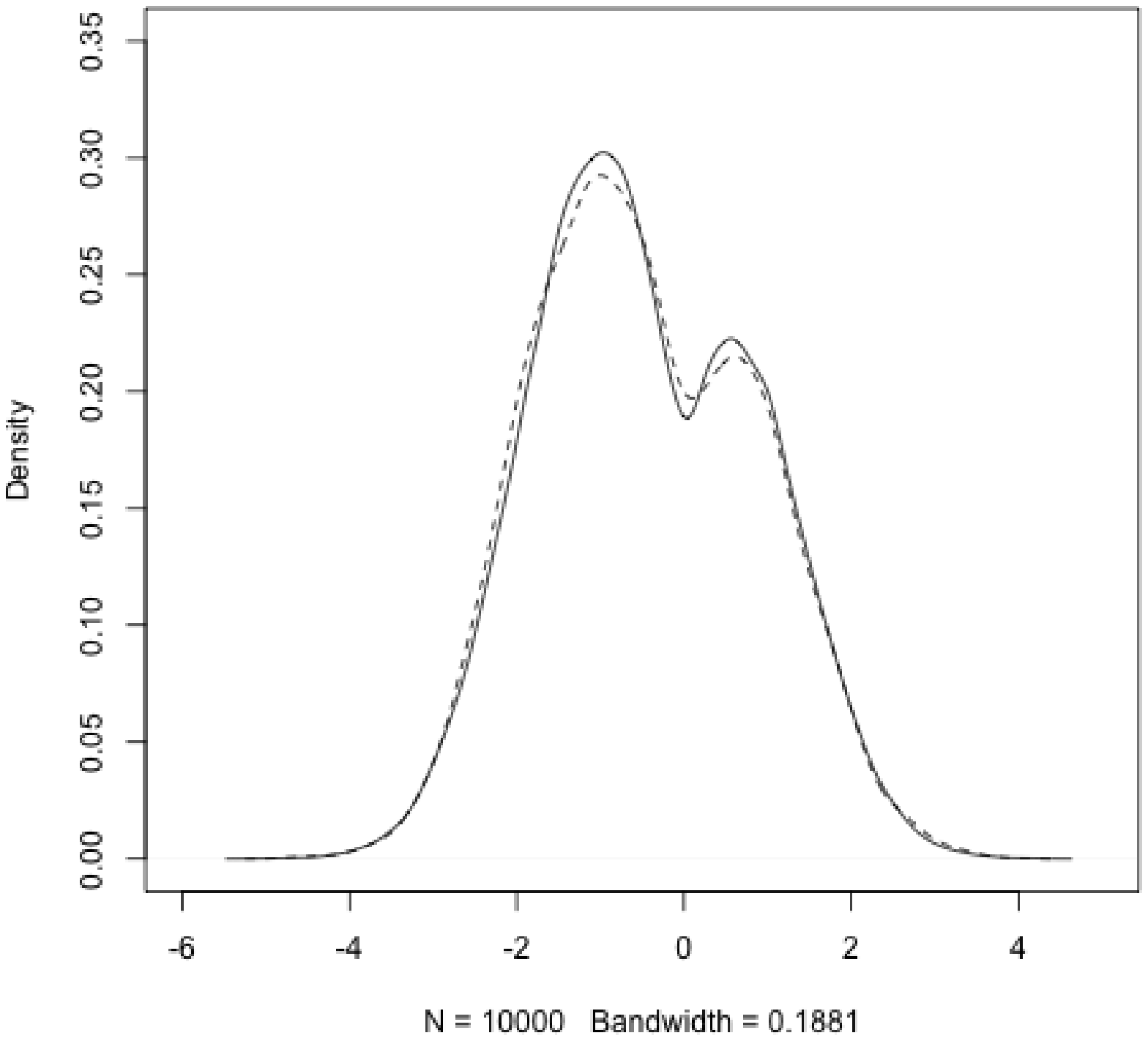}
		\caption*{SDE$(1), a_{1}=0.2, a_{2}= -0.9$}\label{Picture1}		
	\end{subfigure}
	\hspace{0.15cm}
	\begin{subfigure}[t]{5.5cm}
		\centering
		\includegraphics[width=\textwidth]{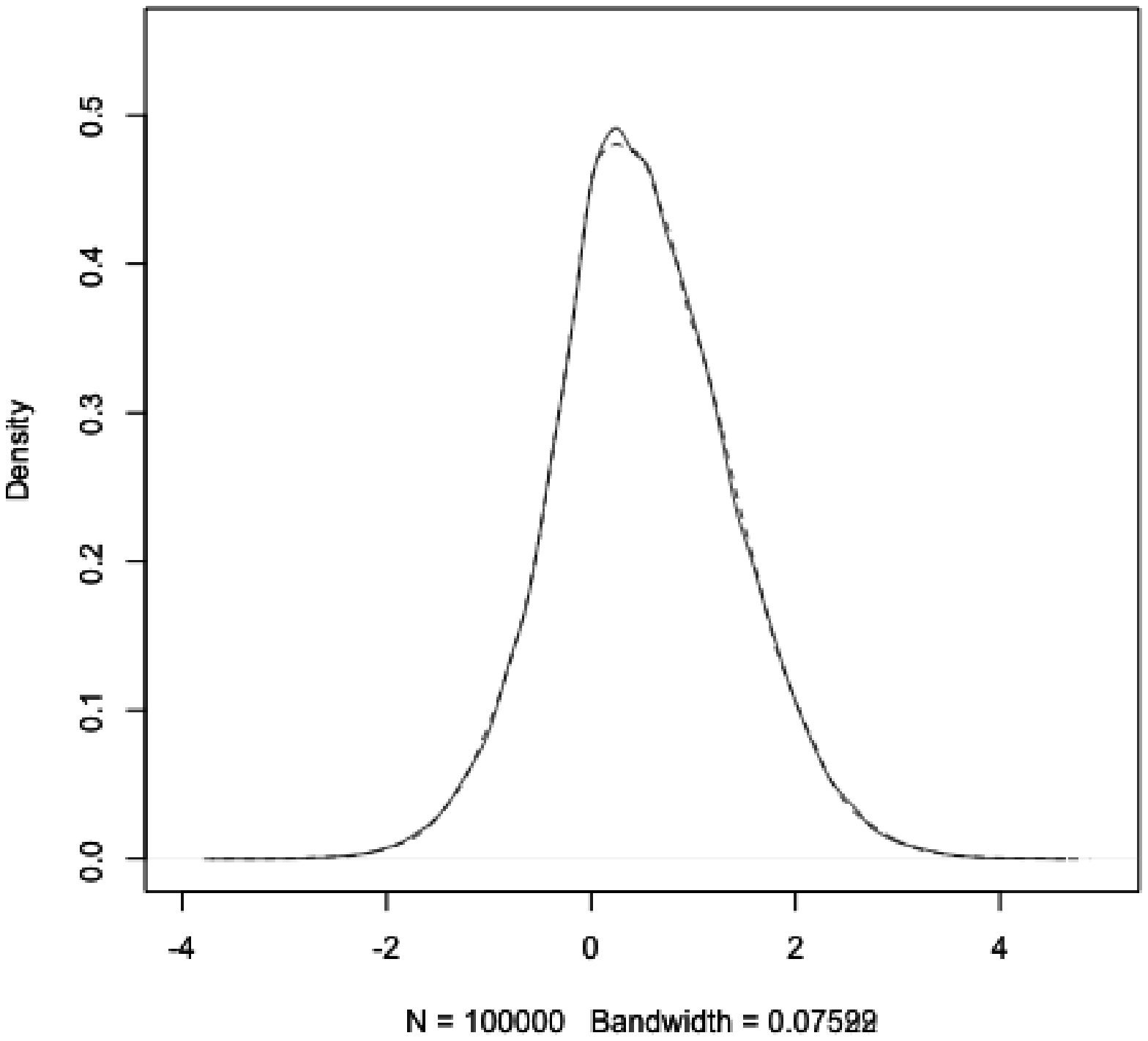}
		\caption*{\hspace{0.3cm} SDE$(1), a_{1}=0.3,a_{2}=0.9$}\label{Picture2}
	\end{subfigure}
\end{figure*}
\begin{figure}
	\begin{center}
	\begin{subfigure}[t]{5.5cm}
		\centering
		\includegraphics[scale=0.35]{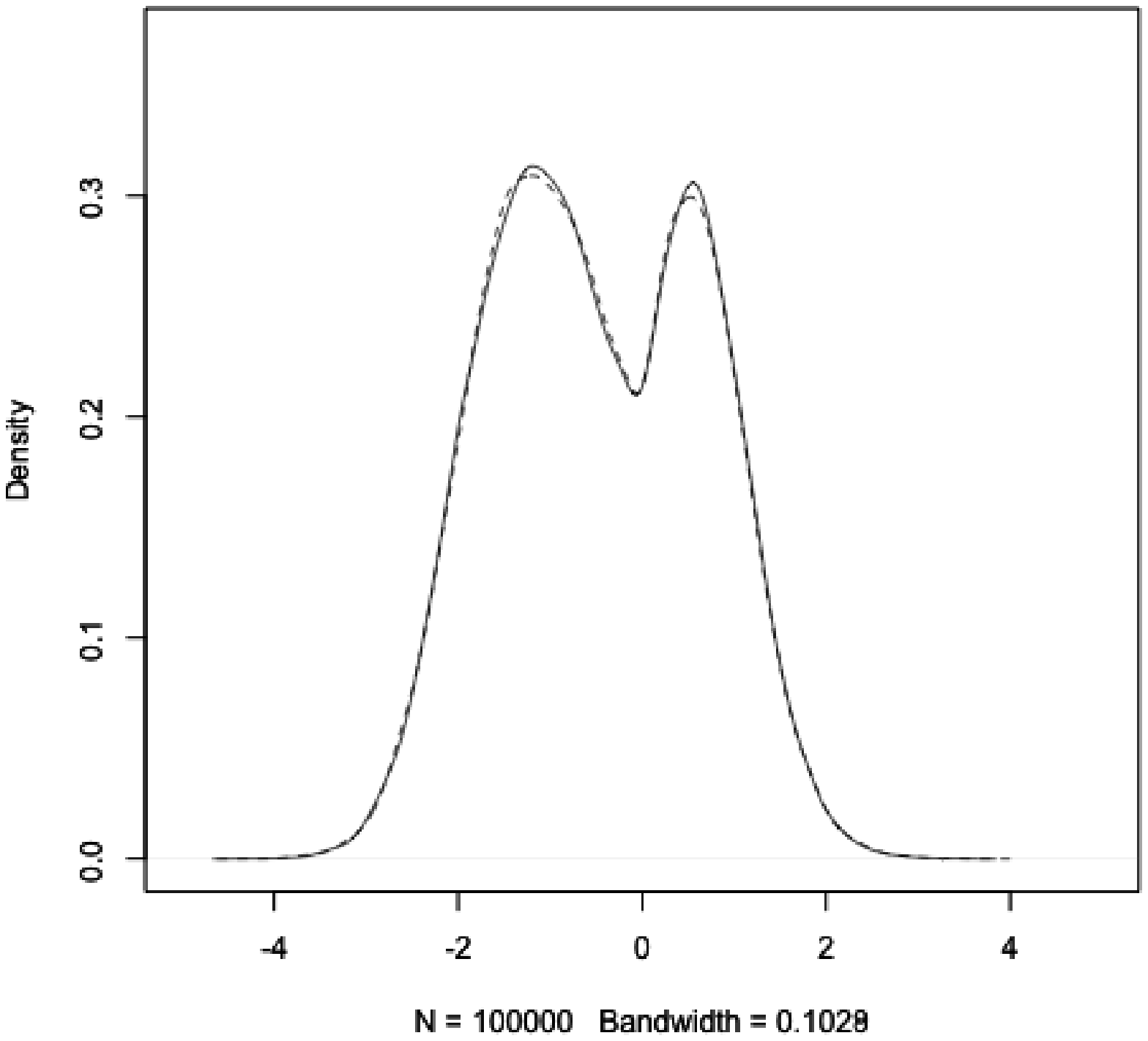}
		\caption*{SDE (2), $\theta_{1}={7 \over 6}\pi, \theta_{2}= {1 \over 4}\pi$}\label{Picture3}		
	\end{subfigure}
	\end{center}
	\caption{Kernel density estimation for $X$ at time $T=1$ using observations obtained by exact methods (solid line) and the Euler-Maruyama method (dashed line).}\label{fig:1}
\end{figure}

\newpage
\section*{Appendix}
Explicit formulas for $p_{1}(l_{1})$ and $p_{3}(l_{3})$ used in Section \ref{sec:double:bridge}.
\begin{align*}
p_{1}(l_{1})=
& \mathbbm{1}_{b_{1} >0} c_{1}e^{ - {  (b_{1}+k_{1})^{2}\over 2(s_{3}-s_{1})}   }   [  \sigma^{2}e^{ -     {\mu_{1}^{2}  \over 2\sigma^{2} }    } +\sqrt{2\pi}\sigma(\mu_{1}+k_{1})   (1- \Phi_{\mu_{1},\sigma^{2}}(0)       )   \\
&  - e^{    { \mu_{2}^{2} -\mu_{1}^{2}  \over 2 \sigma^{2}    }     } (    \sigma^{2}e^{ -     {\mu_{2}^{2}  \over 2\sigma^{2} }    } +\sqrt{2\pi}\sigma(\mu_{2}+k_{1})    (1- \Phi_{\mu_{2},\sigma^{2}}(0)) )]\\
&+ \mathbbm{1}_{b_{1} <0} c_{1}e^{ - {  (b_{1}-k_{1})^{2}\over 2(s_{3}-s_{1})}   }   [  \sigma^{2}e^{ -     {\mu_{3}^{2}  \over 2\sigma^{2} }    } -\sqrt{2\pi}\sigma(\mu_{3}-k_{1})    \Phi_{\mu_{3},\sigma^{2}}(0)          \\
&  + e^{    { \mu_{4}^{2} -\mu_{3}^{2}  \over 2 \sigma^{2}    }     } (   - \sigma^{2}e^{ -     {\mu_{4}^{2}  \over 2\sigma^{2} }    }+\sqrt{2\pi}\sigma(\mu_{4}-k_{1})    \Phi_{\mu_{4},\sigma^{2}}(0) )]
\end{align*}
\begin{align*}
p_{3}(l_{3})
=&\mathbbm{1}_{b_{3} >0} c_{2} e^{ - {  (b_{3}+k_{2})^{2}\over 2(s_{3}-s_{1})}   }  [ \sigma^{2}e^{ -     {\nu_{1}^{2}  \over 2\sigma^{2} }    }  +\sqrt{2\pi}\sigma(\nu_{1}+k_{2})   (1- \Phi_{\nu_{1},\sigma^{2}}(0)       )   \\
&  - e^{    { \nu_{2}^{2} -\nu_{1}^{2}  \over 2 \sigma^{2}    }     } (    \sigma^{2}e^{ -     {\nu_{2}^{2}  \over 2\sigma^{2} }    } +\sqrt{2\pi}\sigma(\nu_{2}+k_{2})    (1- \Phi_{\nu_{2},\sigma^{2}}(0)) )]\\
&+ \mathbbm{1}_{b_{3} <0} c_{2}e^{ - {  (b_{3}-k_{2})^{2}\over 2(s_{3}-s_{1})}   }   [  \sigma^{2}e^{ -     {\nu_{3}^{2}  \over 2\sigma^{2} }    } -\sqrt{2\pi}\sigma(\nu_{3}-k_{2})    \Phi_{\nu_{3},\sigma^{2}}(0)        \\
&  + e^{    { \nu_{4}^{2} -\nu_{3}^{2}  \over 2 \sigma^{2}    }     } (   - \sigma^{2}e^{ -     {\nu_{4}^{2}  \over 2\sigma^{2} }    }+\sqrt{2\pi}\sigma(\nu_{4}-k_{2})    \Phi_{\nu_{4},\sigma^{2}}(0) )]
\end{align*}
\vspace{-0.5cm}
where 
\begin{align*}
 c_{1}&= ( f_{s_{3}-s_{1}}^{b_{1}}(b_{3},l_{3}-l_{1}))^{-1} (2\pi)^{-1}(s_{2}-s_{1})^{-{1 \over 2}}(s_{3}-s_{2})^{- {3 \over 2}}\\
c_{2}&=( f_{s_{3}-s_{1}}^{b_{1}}(b_{3},l_{3}-l_{1}))^{-1} (2\pi)^{-1}(s_{2}-s_{1})^{- {3 \over 2}} (s_{3}-s_{2})^{-{1 \over 2}}\\
k_{1}&=  l_{3}-l_{1}+|b_{3}| \hspace{3cm} k_{2}= l_{3}-l_{1}+|b_{1}|\\
 \mu_{1}&=  { b_{1}(s_{3}-s_{2})-k_{1}(s_{2}-s_{1})   \over s_{3}-s_{1}} \hspace{1cm}  \mu_{2}= \mu_{1}-2b_{1}{s_{3}-s_{2} \over s_{3}-s_{1}}\\
 \mu_{3}&= { b_{1}(s_{3}-s_{2})+k_{1}(s_{2}-s_{1})  \over s_{3}-s_{1}} \hspace{1cm} \mu_{4}= \mu_{3}-2b_{1}{s_{3}-s_{2} \over s_{3}-s_{1}}\\
 \nu_{1}&=  { b_{3}(s_{2}-s_{1})-k_{2}(s_{3}-s_{2})   \over s_{3}-s_{1} } \hspace{1cm}  \nu_{2}= \nu_{1}-2b_{3}{s_{2}-s_{1} \over s_{3}-s_{1}}\\
\nu_{3}&=  { b_{3}(s_{2}-s_{1})+k_{2}(s_{3}-s_{2})   \over s_{3}-s_{1} } \hspace{1cm} \nu_{4}= \nu_{3}-2b_{3}{s_{2}-s_{1} \over s_{3}-s_{1}}\\
\end{align*}
\vspace{-1cm}
$$\sigma^{2}=  (s_{2}-s_{1})(s_{3}-s_{2})( s_{3}-s_{1})^{-1}  $$

\vspace{-0.5cm}

\bibliographystyle{alpha} 
\bibliography{references}

\end{document}